\documentclass[fleqn,10pt]{wlscirep}
\usepackage{geometry}
\usepackage{etoolbox}
\usepackage{graphicx}
\usepackage{color}
\usepackage{graphicx}
\usepackage{amsmath}
\usepackage{capt-of}
\usepackage{caption}
\usepackage{slashbox}

\title{Floquet engineering of topological metal states and hybridization of edge states with bulk states in dimerized two-leg ladders}

\author[1]{Milad Jangjan}
\author[1,*]{Mir Vahid Hosseini}
\affil[1]{Department of Physics, Faculty of Science, University of Zanjan, Zanjan 45371-38791, Iran}
\affil[*]{Corresponding author: mv.hosseini@znu.ac.ir}

\begin{abstract}
We consider asymmetric and symmetric dimerized two-leg ladders, comprising of four different lattice points per unit cell, illuminated by circularly polarized light. In the asymmetric dimerized ladder case, rungs are not perpendicular to the ladder's legs whereas the rungs are perpendicular to the legs for the symmetric one. Using the Floquet theory, we obtain an effective Hamiltonian to study topological properties of the systems. Depending on the dimerization strength and driving amplitude, it is shown that topologically protected edge states manifest themselves not only as a zero-energy band within the gap between conduction and valence band but also as finite-energy curved bands inside the gap of subbands. The latter one can penetrate into bulk states and hybridize with the bulk states revealing hybridized Floquet topological metal phase with delocalized edge states in the asymmetric ladder case. However, in the symmetric ladder, the finite-energy edge states while remaining localized can coexist with the extended bulk states manifesting Floquet topological metal phase.
\end{abstract}
\begin{document}
\flushbottom
\maketitle

\thispagestyle{fancy}

Topological states of matter with intriguing properties have attracted a lot of attention in various fields of physics, particularly, solid-state physics \cite{TopoSolid}. Because of robustness of such states against ubiquitous perturbations \cite{TopoSymmetry}, materials hosting topological states will be excellent candidates for sensitive electronic applications. Topological insulators \cite{TI} along with topological superconductors \cite{Ts} exhibiting topologically nontrivial phases have been interesting topics from theoretical and experimental view points. However, the known topological systems in the equilibrium situation which can indeed be used to realistic applications are limited to a few cases leading to exploring topological quantum states out-of-equilibrium \cite{QuanEngin}.

Beside materials including static topological phases, engineering of exotic nontrivial phases of quantum materials \cite{QuanEngin1} has been developed by means of externally applied dynamical fields. Such approach provides a flexible and practical way to produce desired phases which are absent in the static counterparts. For instance, periodic driving establishes dynamical topological states, known as topological Floquet states \cite{TopoFloquet,TopoFloquet0,TopoFloquet1}. An interesting characteristic of the Floquet theory \cite{Floquet1,Floquet2} is to add extra dimension in a quantum system through continuous evolution over all times within the driving period \cite{micromotion1,micromotion2,TopoFloquet1} providing higher-dimensional systems effectively. In the opposite limit, i.e., stroboscopic picture  \cite{QuanEngin1,TopoFloquet0}, periodic driving manipulates the system parameters expanding phase diagram to values that are not easily accessible in undriven systems. Both of these two features pave the way to turn trivial phases of the system into exotic ones, such as Floquet topological semimetals \cite{TopoFloquetSemi1,TopoFloquetSemi3}, Floquet topological superconductors, \cite{TopoFloquetSuper,TopoFloquetSuper1}, and Floquet topological insulators \cite{TopoFloquet,TopoFloquetInsula}.

There are a variety of techniques for exerting time periodicity and establishing topologically protected edge states such as shining a matter with light \cite{TopoFloquet,Topolight1,Topolight2,Topolight3}, shaking optical lattices \cite{TopoShaking,TopoShaking1} as well as photonic set-ups \cite{TopoPhoto,TopoPhoto1}. Notice, however, that since periodic driving influences on the band structure of system, real space dimensions of the undriven system play significant roles as a basic platform \cite{TopoTable,TopoTable2}. Among the studies of Floquet topological states, one-dimensional (1D) systems have been the center of attention due to the existence of simple models, for example, Su-Schrieffer-Heeger (SSH) model \cite{SSH,SSH1} 
and its generalized versions \cite{GeSSH,GeSSH0,SSHZeeman,GeSSH1}. These models may be served as a building-block for topological quantum information technology \cite{QuantComput2,QuantComput3} owing to supporting non-Abelian statistics. It has been shown that periodically driven 1D systems \cite{TopoTable1,FT1D1} reveal rich Floquet topological features in both the high-frequency \cite{FT1DhighFreq,FT1DhighFreq2} and low-frequency \cite{lowfre} regimes. Moreover, several schemes exploring topologically nontrivial phases with ladder geometry have been proposed in quasi-1D static \cite{ladder,ladder4,ladder5,ladder6} and Floquet \cite{Floquladder0,Floquladder1,Floquladder} systems. Most of studied Floquet topological systems are devoted to cases hosting Floquet topological insulators and superconductors with two-band model. So, in low dimensional quantum systems, it is interesting to extend Floquet topological states, being neither Floquet topological insulators nor Floquet topological superconductors, containing more than two bands \cite{FTfourBand,FTfourBand1} with new exotic topological phases.

In this paper, within the tight-binding approach, we investigate Floquet engineering of topological features of a two-leg SSH chain irradiated by circularly polarized electromagnetic field. We focus on the role of ladder dimerization and geometrical structure as well as driving amplitude to manipulate topological phases of the system using the Floquet formalism. Specifically, we explore topological features for two different situations. Firstly, when the pattern of dimerizations and lattice spacings of the two legs are opposite resulting in asymmetric ladder. Secondly, when the pattern of dimerizations and the lattice spacings of the two legs are identical presenting symmetric ladder. We find that zero-energy flat band and finite-energy band \cite{FTfourBand,FiniteEneFlat2} can  be emerged in the asymmetric ladder case. Through such a simple class of model, we further, interestingly, find that these finite-energy edge states can reside within bulk states and would hybridize with them resulting in hybridized Floquet topological metal phase. However, for symmetric ladder model, the zero-energy edge states disappear and also the hybridization of finite-energy edge states with the bulk ones suppresses while the edge states are within the bulk states giving rise Floquet topological metal phase. Unlike the previous finite-energy flat band cases \cite{FTfourBand,FiniteEneFlat2}, whose quasi-energies are fixed at the edge of Floquet zone, here, such midgap states which are not necessarily flat occur in subband gaps and their energy values can be adjusted by dimerization strength and driving amplitude.

This paper is organized as follows. In Sec. 2, we introduce our model and its Hamiltonian. The conditions of band touching points are derived. The relevant topological invariants associated with the existing symmetries of the system are discussed in Sec. 3. The analysis of topological features of asymmetric and symmetric ladder, respectively, is given in Secs. 4 and 5. Also, the stability of topological phases against symmetry breaking perturbations is investigated in Sec. 6. Finally, we conclude with a summary and discussion in Sec. 7.

\section*{Model and Theory}\label{s2}

\begin{figure}[t!]
\centerline{\includegraphics[width=17cm]{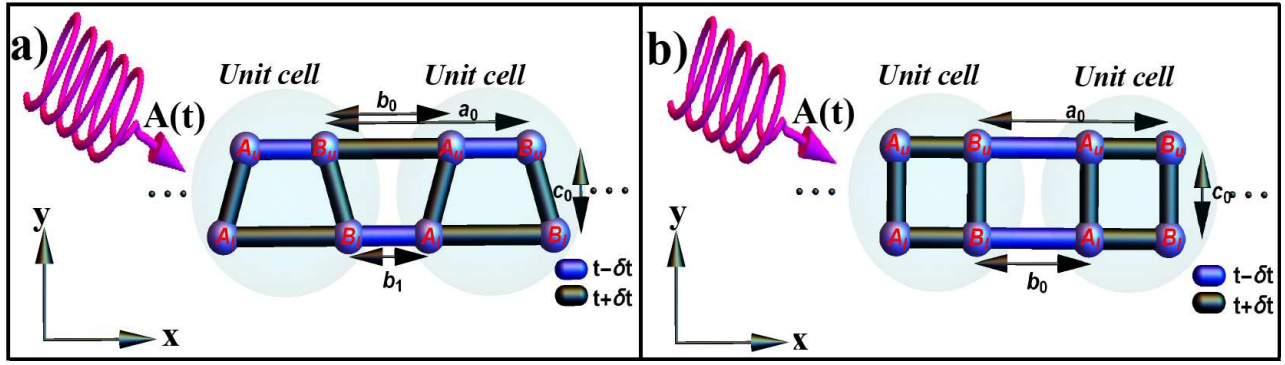}}
\caption{(Color online) Two-leg ladder that each leg describes SSH chain under the light irradiation. (a) Asymmetric ladder geometry with opposite dimerization of the legs and different inter unit cell spacings $b_0$ and $b_1$ of the upper and lower leg, respectively. (b) Symmetric ladder geometry with identical dimerization and inter unit cell spacing $b_0$ of the legs. $a_0$ is the length of unit cell and $c_0$ is the inter chain distance.}
\label{fig1}
\end{figure}

We consider a system consisting of a two-leg ladder that each leg describes SSH chain in the presence of light illumination, as represented in Fig. \ref{fig1}. We will examine topological properties for two different cases; (i) asymmetric ladder case where the dimerization and the corresponding lattice spacings of legs ($b_0 \ne b_1$) are asymmetric [see Fig. \ref{fig1}(a)], and (ii) symmetric ladder case where the dimerization of both SSH chains is identical and the corresponding lattice spacing of legs is equal ($b_0=b_1$) [see Fig. \ref{fig1}(b)]. In the absence of irradiation, the tight-binding Hamiltonian of this model containing four sublattices per unit cell can be written as
\begin{align}\label{e1}
H=\sum_j^N [t_1 A^{\dagger}_{uj}B_{uj} +t_2 A^{\dagger}_{lj}B_{lj} + t_3 A^{\dagger}_{uj} A_{lj} + t_4 B^{\dagger}_{uj}B_{lj}] + \sum_j^{N-1} [t_1^\prime B^{\dagger}_{uj}A_{u{j+1}}+t_2^\prime B^{\dagger}_{lj}A_{l{j+1}}] + h.c,
\end{align}
where $X^{(\dagger)}_{u/lj}$ is the electron annihilation (creation) operator of sublattice $X$ (which can be either A or B type) on the upper/lower chain at the $j$th unit cell. $t_1^{(\prime)}$ and $t_2^{(\prime)}$ are intra (inter) unit cell hoppings along the upper and lower legs, respectively. The hopping energies along the rungs of the ladder are $t_3$ and $t_4$. We choose $t_1=t_2^\prime=t-\delta t$ and $t_2=t_1^\prime=t+\delta t$ for asymmetric ladder whereas $t_1=t_2=t+\delta t$ and $t_1^\prime=t_2^\prime=t-\delta t$ for symmetric ladder where $\delta t=\delta_0\cos\theta$ is the dimerization strength with $\theta$ and $\delta_0$ being a cyclical parameter varying from $0$ to $2\pi$ continuously and dimerization amplitude, respectively. For both asymmetric and symmetric cases, we choose $t_3=t_4=t+\delta t$. Notably, the symmetric ladder relies on poly acetylene including identical dimerization of chains. We also set $t$ as a unit of energy, the lattice constant $a_0$ as a length unit. Throughout the paper $\delta_0$= 0.8 without loss of generality.

In the presence of externally applied electromagnetic field comprising of the periodic time-dependent electric field $\mathbf{E}(t)=-\partial_t\mathbf{A}(t)$ with vector potential
\begin{align}\label{vec}
\mathbf{A}(t) = (A_xsin\omega t, A_ysin (\omega t +\phi)),
\end{align}
Hamiltonian (\ref{e1}) can be periodic in time
$H(t)=H(t+T)$ through Peierls substitution
\begin{align}\label{Peierls}
t_{ij} \longrightarrow t_{ij}e^{-\frac{ie}{\hslash c}\int_{R_{i}}^{R_{j}}\mathbf{A}(t)\cdot d\mathbf{r}}.
\end{align}
Here, $A_{x(y)}=\frac{E_{x(y)}}{\omega}$ is the driving amplitude along the x (y)-direction which can be related to the amplitude of electric field $E_{x(y)}$. The period $T=\frac{2\pi}{\omega}$ is determined by the driving frequency $\omega$, $\phi$ is a phase shift, c is light speed, and $e$ is electron charge. We take $\hslash=1$ and $\frac{e}{c}=1$ hereafter.

Floquet theorem \cite{Floquet1,Floquet2} can be used to find a solution to the time-dependent Schrodinger equation with time-periodic potentials. This theorem guarantees the existence of a set of solutions
\begin{equation}\label{e2}
\psi_n(t)=e^{-i\epsilon_n t} \varphi_n (t),
\end{equation}
where $\epsilon_n$ is the Floquet quasi-energy and Floquet state $\varphi_n (t)$ has the same time periodicity as Hamiltonian, $\varphi_n(t+T)=\varphi_n (t)$, in analogy with Bloch theorem in which the so-called Bloch states are periodic in real space.
For every solution $\varphi_n (t)$ with quasi-energy $\epsilon_n$ one can construct another solution $\varphi_{\alpha n}(t)=exp(-i\alpha \omega t) \varphi_n(t)$ with quasi-energy $\epsilon_{n\alpha}=\epsilon_{n}+\alpha\omega$, that corresponds to the same physical state $\varphi_n(t)$.
In fact, the Floquet states are the solutions of the eigenvalue equation
\begin{equation}\label{e3}
 H_F\vert \varphi_n(t) \rangle=\epsilon_n\vert \varphi_n(t)\rangle,
\end{equation}
where $H_F=H-i\frac{\partial }{\partial t}$ is Floquet Hamiltonian. Eventually, matrix elements of the Floquet Hamiltonian can be written as,
\begin{equation}\label{e4}
H_F^{\alpha\beta}=\frac{1}{T}\int _{0}^{T} H(t) e^{i(\alpha-\beta)\omega t}dt - \alpha \omega \delta_{\alpha \beta},
\end{equation}
where $\alpha$ and $\beta$ are Floquet index.
Hence, by involving Peierls substitution (\ref{Peierls}) in the static Hamiltonian (\ref{e1}) and using Eq. (\ref{e4}), the Floquet Hamiltonian can be obtained as
\begin{align}\label{e5}
H_F^{\alpha\beta} =\sum_j^N [\tilde{t}_1^{\alpha\beta}  A^{\dagger}_{uj}B_{uj} + \tilde{t}_2^{\alpha\beta} A^{\dagger}_{lj}B_{lj}
+\tilde{t}_3^{\alpha\beta}  A^{\dagger}_{uj} A_{lj} + \tilde{t}_4^{\alpha\beta} B^{\dagger}_{uj}B_{lj}]+ \sum_j^{N-1} [\tilde{t^{\prime}}_1^{\alpha\beta} B^{\dagger}_{uj}A_{u{j+1}} + \tilde{t^{\prime}}^{\alpha\beta}_2 B^{\dagger}_{lj}A_{l{j+1}}] + h.c - \alpha \omega \delta_{\alpha \beta}.
\end{align}
Here, we have defined
\begin{eqnarray}\label{e6}
\tilde{t}_1^{\alpha\beta}&=&t_1J_{\alpha-\beta}[A_{x}(a_0-b_0)], \nonumber \\
\tilde{t}_2^{\alpha\beta}&=&t_2J_{\alpha-\beta}[A_{x}(a_0-b_1)], \nonumber \\
\tilde{t}_3^{\alpha\beta}&=&t_3J_{\alpha-\beta}[\sqrt{(A_xb_2)^2+(A_yc_0)^2+2A_xb_2A_yc_0\cos\phi}], \nonumber \\
\tilde{t}_4^{\alpha\beta}&=&t_4J_{\alpha-\beta}[\sqrt{(A_xb_2)^2+(A_yc_0)^2-2A_xb_2A_yc_0\cos\phi}], \nonumber \\
 \tilde{t^{\prime}}_1^{\alpha\beta}&=&t^{\prime}_1J_{\alpha-\beta}[A_{x}b_0],\nonumber \\
 \tilde{t^{\prime}}_2^{\alpha\beta}&=&t^{\prime}_2J_{\alpha-\beta}[A_{x}b_1],
\end{eqnarray}
where $b_2=(b_0-b_1)/2$ and $J_m[x]$ is the first kind Bessel function of order $m$. Considering the high-frequency regime (off-resonant regime) where the Floquet bands are decoupled from each other, the system can be well described by zeroth order static Floquet Hamiltonian
\begin{eqnarray}\label{e7}
 H_F^{00} =\sum_j^N [\tilde{t}_1^{00}  A^{\dagger}_{uj}B_{uj} +\tilde{t}_2^{00}A^{\dagger}_{lj}B_{lj}+\tilde{t}_3^{00}  A^{\dagger}_{uj} A_{lj}+ \tilde{t}_4^{00} B^{\dagger}_{uj}B_{lj}]+ \sum_j^{N-1} [\tilde{t^{\prime}}^{00}_1 B^{\dagger}_{uj}A_{u{j+1}} +\tilde{t^{\prime}}^{00}_2 B^{\dagger}_{lj}A_{l{j+1}}] + h.c.
\end{eqnarray}
In the following, we omit the super index "$00$" from the parameters of Eq. (\ref{e7}) for the sake of brevity.

To study the bulk properties of system, we impose the periodic boundary conditions and take Fourier transformation  $X_{u/lj}=\frac{1}{\sqrt{N}}\Sigma_k e^{-ikj} X_{u/lk}$, where N is the number of the unit cells. Then the Hamiltonian can be written in the form of
\begin{equation}\label{e8}
H_F=\sum_k\psi^\dagger_k h_F(k)\psi_k,
\end{equation}
where $\psi^\dagger_k=(A_{uk},A_{lk},B_{lk},B_{uk})^\dagger$ and
\begin{eqnarray}\label{e9}
 h_F(k)=
 \left ( \begin{array}{c c c c}
     0 & \tilde{t}_3 & 0& \tilde{t}_1+\tilde{t}_1^\prime e^{ik} \\
     \tilde{t}_3 & 0 & \tilde{t}_2+\tilde{t}_2^\prime e^{ik} & 0 \\
      0 & \tilde{t}_2+\tilde{t}_2^\prime e^{-ik} & 0 & \tilde{t}_4 \\
     \tilde{t}_1+\tilde{t}_1^\prime e^{-ik} & 0 &\tilde{t}_4 & 0
 \end{array}\right).\nonumber \\
\end{eqnarray}
After diagonalizing the Hamiltonian (\ref{e9}), the eigenvalues can be obtained in the momentum space as
\begin{equation}\label{e10}
E_{l,p}(k)=\frac{l \sqrt{\zeta +p \sqrt{\eta}}}{\sqrt{2}},
\end{equation}
with
\begin{eqnarray}\label{e11}
\zeta &=& \tilde{t}_3^2 + \tilde{t}_4^2+\tilde{t}_1^2 + \tilde{t}_2^2+ \tilde{t}_1^\prime+\tilde{t}_2^\prime + 2(\tilde{t}_1\tilde{t}_1^\prime +\tilde{t}_2\tilde{t}_2^\prime)cos(k), \nonumber \\
\eta &=& \zeta^2-4(\tilde{t}_2^2\tilde{t}_1^{\prime2}+(\tilde{t}_3\tilde{t}_4-\tilde{t}_1^\prime\tilde{t}_2^\prime)^2+2\tilde{t}_1 \tilde{t}_2(-\tilde{t}_3 \tilde{t}_4+\tilde{t}_1^\prime\tilde{t}_2^\prime)+\tilde{t}_1^2(\tilde{t}_2^2 +\tilde{t}_2^{\prime2})+
 2(\tilde{t}_1 \tilde{t}_2^\prime + \tilde{t}_2 \tilde{t}_1^\prime)(\tilde{t}_1 \tilde{t}_2 - \tilde{t}_3 \tilde{t}_4 + \tilde{t}_1^\prime \tilde{t}_2^\prime)cos(k) \nonumber \\
 &+&2\tilde{t}_1 \tilde{t}_2 \tilde{t}_1^\prime \tilde{t}_2^\prime cos(2k)),
\end{eqnarray}
where the band index $l=-(+)$ stands for valence (conduction) band and $ p=+(-) $ indicates upper (lower) subband. The topological phase transition is accompanied by closing and reopening the gap at the super-symmetry points of k-space, i.e., $k=0$ and $k=\pi$. It is straightforward to see that the conditions of gap closing between the two valence bands can be obtained by solving $E_{l=-,p=-}(k)=E_{l=-,p=+}$, yielding
\begin{eqnarray}\label{e12}
t_a+e^{ik}t_a^\prime &=&\pm \sqrt{-(\tilde{t}_3-\tilde{t}_4)^2}, \ \textrm{if}\ \tilde{t}_3 = \tilde{t}_4, \nonumber \\
 t_b+e^{ik}t_b^\prime &=& \pm \sqrt{-(\tilde{t}_3+\tilde{t}_4)^2}, \ \textrm{if}\ \tilde{t}_3 =-\tilde{t}_4,
\end{eqnarray}
at the momentum $k=0$ and $k=\pi$. Here, we have defined $t_a=\tilde{t}_1+\tilde{t}_2$, $t_a^\prime =\tilde{t}_1 ^\prime + \tilde{t}_2^\prime$, $t_b=\tilde{t}_1-\tilde{t}_2$, and $t_b^\prime = \tilde{t}_1^\prime - \tilde{t}_2^\prime$. As can be seen from above equations, the square root expression must be zero to occur topological phase transition. Also, the gap closure conditions between the upper valence band ($l=-,p=+$) and lower conduction band ($l=+,p=-$) are
\begin{eqnarray}\label{e14}
(t_a+ e^{ik} t_a^\prime)^2-(t_b+e^{ik}t_b^\prime)^2 =4\tilde{t}_3\tilde{t}_4,
\end{eqnarray}
at the momentum $k=0$ and $k=\pi$. Equations (\ref{e12}) and (\ref{e14}) represent boundaries between topologically distinct phases where the value of topological invariant will be changed at these points.

We define exchange operator $\Upsilon$ that exchanges the two legs of ladder and their corresponding sublattices as
\begin{eqnarray}\label{e17}
\Upsilon \psi \rightarrow \psi '=\left(\begin{array}{c}
A_{l}\\
A_{u}\\
B_{u}\\
B_{l}
\end{array}\right).
\end{eqnarray}
In the basis of exchange operator, obviously, $\Upsilon$ must be diagonalized,
\begin{eqnarray}
U_1\Upsilon U_1^{-1}= \left(\begin{array}{c c c c}
-1&0&0&0 \\
0&-1&0&0 \\
0&0&1&0 \\
0&0&0&1
\end{array}\right),
\end{eqnarray}
through the unitary matrix
\begin{eqnarray}\label{e027}
U_1&=&\frac{1}{\sqrt{2}}\left(\begin{array}{c c c c}
0&0&-1&1 \\
-1&1&0&0 \\
0&0&1&1 \\
1&1&0&0
\end{array}\right).
\end{eqnarray}
Transforming Hamiltonian (\ref{e9}), with the unitary matrix $U_1$, yields
\begin{eqnarray}\label{e26}
\tilde{h}_F=U_1 h_F(k)U_1^{-1}=\left(\begin{array}{c c}
 h_{1} & h_{cou} \\
-h_{cou}^ \star & -h_1
\end{array}\right),
\end{eqnarray}
where
\begin{eqnarray}\label{e27}
h_1&=&\frac{1}{2}\left(\begin{array}{c c}
 \tilde{t}_4&t_a+t_a ^\prime e^{ik}\\
t_a+t_a ^\prime e^{-ik} & \tilde{t}_3
\end{array}\right),  \nonumber \\
h_{cou}&=&\frac{1}{2}\left(\begin{array}{c c}
 0&t_b+t_b ^\prime e^{ik}\\
-t_b-t_b ^\prime e^{-ik} &0
\end{array}\right).
\end{eqnarray}
From Hamiltonian (\ref{e26}), one finds that the diagonal blocks ($h_1$, $-h_1$) are the well-studied Hamiltonian of generalized SSH model \cite{GeSSH} which are coupled by the off-diagonal block $h_{cou}$. Note, the structure of matrix (\ref{e26}) implies that the energy spectra of the individual diagonal block will be shifted from zero energy and the off-diagonal block $h_{cou}$ is responsible for opening a gap around zero energy. Therefore, one may expect that Hamiltonian (\ref{e26}) has two kinds of edge states, one of them is zero-energy edge states which may be protected by symmetries of the whole Hamiltonian and another is finite-energy edge states due to SSH-like analogue of the block $h_1$ which may be protected by symmetries of the diagonal block.

It is easy to check that Hamiltonian (\ref{e26}) has time-reversal and particle-hole symmetry defined, respectively, as $\mathcal{T} \tilde{h}_F(k)\mathcal{T}=\tilde{h}^\star_F(-k)$ and $\mathcal{P} \tilde{h}_F(k)\mathcal{P}=-\tilde{h}^\star_F(-k)$ with the corresponding operators $\mathcal{T}=\sigma_0 \otimes \sigma_0 \mathcal{K}$ and $\mathcal{P}=\sigma_{x} \otimes \sigma_0 \mathcal{K}$ where $\sigma_{0}$ and $\sigma_x$ being the identity matrix and x component of Pauli matrix. $\mathcal{K}$ is complex conjugate operator. In fact, since $\mathcal{T}\cdot\mathcal{P}=\mathcal{C}$, the unitary chiral operator can be determined as $\mathcal{C}=\sigma_{x} \otimes \sigma_0$. Also, in addition to the mentioned symmetries, under the condition $\tilde{t}_3=\tilde{t}_4$ the Hamiltonian (\ref{e26}) has inversion symmetry with operator $\Pi= \sigma_z \otimes \sigma_x$ as a result of the inversion symmetry of the diagonal blocks.

Before proceeding, to distinguish localized and extend states, we use the logarithm of inverse participation ratio (IPR) which is given by \cite{IPR}
\begin{equation}\label{e25}
 I(E)=\frac{Ln \sum ^{j=4N}_{j=1}|\psi(j)|^4}{Ln4N}.
\end{equation}
Here $\psi (j)$ is the eigenvector at site j with energy $E$. When the IPR is close to zero, the wave function is more localized (energy levels shown in red in the figures). But for extended wave function IPR tends to -1 (energy levels shown in blue in the figures).

\section*{Relevant topological invariants} \label{s3}

The bulk-edge correspondence is a hallmark to confirm the topological feature of system relating topological edge states under open boundary conditions to the bulk topological invariants \cite{TI} calculated under periodic boundary conditions. Therefore, topological invariants of Hamiltonian (\ref{e26}) should predict nontrivial values in the space of parameters where edge states are emerged under open boundary conditions. In the following, we introduce three relevant topological invariants to characterize properly the topology of edge states due to the existence of certain symmetries in the whole and/or diagonal block of Hamiltonian.

First, one of the relevant topological invariants is $\mathcal{Z}$ \cite{Ztopo} that originates from the inversion symmetry of the diagonal blocks, i.e., ($h_1, -h_1$). Each of the diagonal blocks can commute with the inversion operator at the super symmetry points $k=0$ and $\pi$. Hence, the eigenstates of $h_1$ have a well-defined parity at supersymmetry points. Subsequently, one can define an integer invariant for each band gap of the system as
\begin{equation}\label{e30}
    \mathcal{N}_{i,j}=|\mathcal{E}_{1,i,j}-\mathcal{E}_{2,i,j}|,
\end{equation}
where $\mathcal{E}_{1,i,j}$ and $\mathcal{E}_{2,i,j}$ are the number of negative parities of band structure, respectively, at the $k=0$ and $k=\pi$ in the $i$th bandgap of $j$th subspace. Eventually, by using the relation \cite{Tm2}
\begin{equation}\label{e31}
    \mathcal{Z}:=\sum_j \sum_i \mathcal{N}_{i,j},
\end{equation}
we can expose the topology of finite-energy edge states, originating from the diagonal blocks, under open boundary condition.

Second, it is well-known that a relevant topological invariant for quantum system with chiral symmetry which determines topologically distinct phase is winding number. The winding number enumerates the number of pairs of zero-energy edge states. The chiral symmetric Hamiltonian (\ref{e26}) can be brought into a block off-diagonal form in the basis of chiral operator. This can be done by the unitary operator
\begin{equation}\label{e32}
    U_2=\frac{1}{\sqrt{2}}\left(\begin{array}{c c c c}
         0 & -1 & 0 &1 \\
         -1 & 0 & 1 & 0 \\
         0 & 1 & 0 & 1 \\
         1 & 0 & 1 & 0
  \end{array}\right).
\end{equation}
Transforming Hamiltonian (\ref{e26}) by $U_2$ leads to
\begin{equation}\label{e33}
    U_2 \tilde{h}_F(k)
U_2^{-1}=\left(\begin{array}{c c}
    0 & G \\
    G^\dagger & 0
    \end{array}\right),
\end{equation}
where
\begin{equation}\label{e34}
    G=\left(\begin{array}{c c}
    \tilde{t}_4 & \tilde{t}_2+\tilde{t}_2e^{ik} \\
    \tilde{t}_1+\tilde{t}_1e^{-ik} & \tilde{t}_3
    \end{array}\right).
\end{equation}
Now, we can use the following relation to obtain the winding number \cite{Class10,superconduc}
\begin{equation}\label{e35}
    \mathcal{W}=\frac{1}{2\pi i}\int_{-\pi} ^{\pi}dk \partial_k Ln(Z(k)),
\end{equation}
where
\begin{eqnarray}\label{e36}
    Z(k)=Det(G)=\tilde{t}_3\tilde{t}_4-\tilde{t}_1\tilde{t}_2-\tilde{t}_1^\prime\tilde{t}_2^\prime -(\tilde{t}_1\tilde{t}_2^\prime+\tilde{t}_1^\prime\tilde{t}_2)cos(k) -i(\tilde{t}_1\tilde{t}_2^\prime-\tilde{t}_1^\prime\tilde{t}_2)sin(k).
\end{eqnarray}
The integral of Eq. (\ref{e35}) can be evaluated analytically via Cauchy's residue theorem. We find a simple formula characterizing the topology of the system associated with zero-energy edge states as
\begin{eqnarray}\label{e37}
\mathcal{W}= \Theta(x-y)\Theta(x+y)+\Theta(-x+y)\Theta(-x-y),
\end{eqnarray}
where
\begin{eqnarray}\label{e38}
x&=&-(\tilde{t}_1\tilde{t}_2^\prime+\tilde{t}_1^\prime\tilde{t}_2), \nonumber \\
y&=&\tilde{t}_3\tilde{t}_4-\tilde{t}_1\tilde{t}_2-\tilde{t}_1^\prime\tilde{t}_2^\prime,
\end{eqnarray}
and $\Theta(\xi)$ is the Heaviside function. $\mathcal{W}=1$ means the system hosts one pairs of topological edge states at zero energy and $\mathcal{W}=0$ shows trivial topological phase where the system is an ordinary insulator.

Third, when the chiral symmetry is broken by symmetry breaking perturbations, the inversion symmetry of the whole Hamiltonian allows us to use the multi-band Zak phase \cite{ZakPh}
\begin{eqnarray}\label{e39}
\gamma=\sum_{E<0}\int \langle u(k)\vert i\nabla _k \vert u(k) \rangle dk,
\end{eqnarray}
to calculate topological invariant of zero-energy edge states. Here, $\vert u(k)\rangle$ is occupied Bloch states with the corresponding eigenvalue $E$.

\section*{Asymmetric ladder case} \label{s4}

Now, we study band structures and topological properties of asymmetric ladder irradiated by circularly polarized light [see Fig. \ref{fig1}(a)]. We apply the light beam with the vector potential (\ref{vec}) involving circular polarization, i.e., $A_x=A_y=A$. Then the hoppings of Eqs. (\ref{e6}) reduce as
\begin{eqnarray}\label{e60}
\tilde{t}_1&=&t_1J_0[A(a_0-b_0)], \nonumber \\
\tilde{t}_2&=&t_2J_0[A(a_0-b_1)], \nonumber \\
\tilde{t}_3&=&t_3J_0[A\sqrt{b_2^2+c_0^2+2b_2c_0\cos\phi}], \nonumber \\
\tilde{t}_4&=&t_4J_0[A\sqrt{b_2^2+c_0^2-2b_2c_0\cos\phi}], \nonumber \\
 \tilde{t}_1^{\prime}&=&t^{\prime}_1J_0[Ab_0], \nonumber \\
 \tilde{t}_2^{\prime}&=&t^{\prime}_2J_0[Ab_1].
\end{eqnarray}
Note that if $\phi= n\pi/2$ with $n$ an odd number, then the two hoppings of rungs are equal, $\tilde{t}_3 = \tilde{t}_4$. We set $\phi= \pi/2$ in the current section. The case $\phi \ne n\pi/2$ where $\tilde{t}_3 \ne \tilde{t}_4$ owing to $2b_2=b_0 - b_1\ne0$ will be discussed in Sec. 6. Remarkably, in the asymmetric ladder case, the symmetry operators are the same as those for Sec. 2 with features $\mathcal{T}^2=1$, $\mathcal{P}^2=1$, and $\mathcal{C}^2=1$, so the symmetry class belongs to BDI \cite{Class10,AZClass1,AZClass3,AZClass4}. It is worthwhile noting that if we regard the leg degrees of freedom as spin degrees of freedom, then, in the asymmetric ladder, the unequal hopping of upper and lower legs resembles spin-dependent hopping, i.e., spin-orbit interaction. As such, the exchange operator plays the role of spin rotation operator.

\begin{figure*}[th]
  \centering
    \includegraphics[width=1\linewidth]{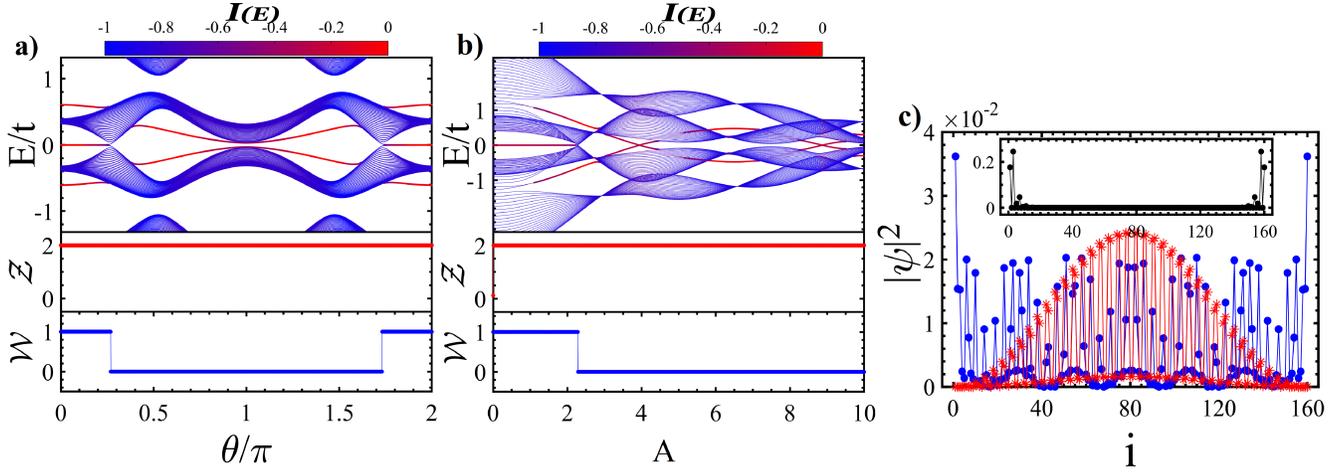}
    \caption{(Color online) Quasi-energy spectrum along with zero- and finite-energy edge states and their relevant topological invariants $\mathcal{Z}$ and $\mathcal{W}$ (a) as a function of  $\theta/\pi$ with $A = 3.2$ and (b) as a function of $A$ with $\theta/\pi$ =0.45. The colors in the energy spectrum represent IPR of the wave function localization. (c) The probability distribution of energy states; Main panel: bulk states (the red curve with aster symbol) and the hybridized edge states with bulk states (the blue curve with circle symbol). Inset: the localized edge states within band gap. Here, $b_0=0.2, b_1=0.1$ and $c_0=0.6$.}
    \label{fig4}
\end{figure*}

Using Eqs. (\ref{e60}), the energy spectra of Hamiltonian (\ref{e7}) can be obtained numerically under open boundary conditions. The dependence of quasi-energy spectra and the appropriate bulk topological invariants on $\theta/\pi$ and on $A$, respectively, is shown in Figs. \ref{fig4}(a) and \ref{fig4}(b). As already predicted above, there exist two kinds of edge states: zero-energy edge states with flat band and finite-energy edge states. As will be shown in Sec. 6, the former can be protected by the chiral or inversion symmetry of the whole Hamiltonian with the corresponding $\mathcal{W}$ or $\gamma$ invariant, respectively. While the latter is protected by the inversion symmetry of block $h_1$ with the corresponding $\mathcal{Z}$ invariant.

From Fig. \ref{fig4}(a) one can see that, interestingly, without occurring topological phase transition, the finite-energy edge states can leave from an energy gap and enter to a new one by passing through bulk states. In such process, the $\mathcal{Z}$ invariant exhibits a nontrivial value resulting in the existence of symmetry protected edge states inside the topological bulk states. Furthermore, the finite-energy edge states hybridize with the extended bulk ones establishing hybridized Floquet topological metal phase with less localized topological edge states. As a result, by varying $\theta$, the values of IPR of edge states change significantly in transition from topological insulator phase, where the edge states are within gapped states, to the hybridized Floquet topological metal phase originating from breaking of the exchange symmetry $\Upsilon$ in the asymmetric ladder case.

Also, as shown in Fig. \ref{fig4}(b) with the increase of driving amplitude $A$ the energies of finite-energy edge states decrease non-monotonically manifesting, alternatively, topological insulator and hybridized Floquet topological metal phases. Furthermore, the zero-energy edge states characterized by the topological invariant $\mathcal{W}$ as functions of $\theta/\pi$ and $A$ reveal either topologically nontrivial stable or trivial phases which are separated by topological phase transition.

To gain insight into the nature of states, in Fig. \ref{fig4}(c), we have plotted the probability distribution of hybridized and localized finite-energy edge states and of the bulk states as a function of unit cell index along the ladder. As usual the localized edge states [see the inset] and the extended bulk states [see the red curve indicated by "aster" symbols in the main panel] have highest probability, respectively, at the ends and in the middle of the system. Moreover, one finds that the hybridized edges states can have finite probability both at the ends and on the bulk of system [see the blue curve marked by "circle" symbols in the main panel].

\begin{figure}[ht!]
\centerline{\includegraphics[width=10.5cm]{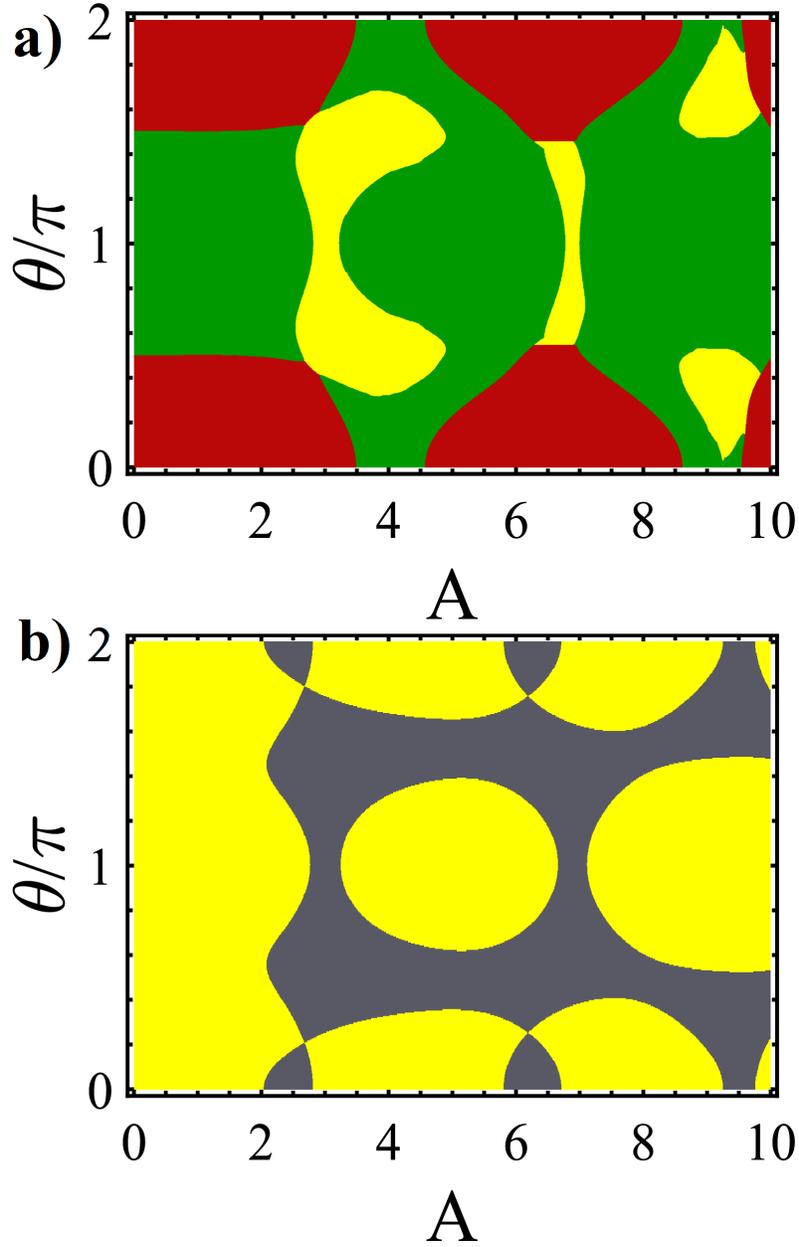}}
\caption{(Color online) Topological phase diagram in ($\theta/\pi$, $A$)-plane associated with (a) finite-energy edge states which the red, green, and yellow regions indicate nontrivial topological regions with related topological invariant $\mathcal{Z}=2$ where the corresponding edge states lie, respectively, within subband gap, within bulk states, and in the main gap and (b) zero-energy edge states which the yellow and the gray regions are related to the topologically non-trivial ($\mathcal{W}=1$) and trivial phase ($\mathcal{W}=0$), respectively. Here, $b_0=0.2, b_1=0.1$, and $c_0=0.6$.}
\label{phase}
\end{figure}

The phase diagrams in the plane ($\theta/\pi$, $A$) including topologically distinct phases with finite- and zero-energy edge states, respectively, are shown in Figs. \ref{phase}(a) and \ref{phase}(b). We represent the topological phases in which the edge states reside in the gap between the subbands and in the main gap by red and yellow colors, respectively. Also, the hybridized Floquet topological metal phase and normal insulator are indicated by green and gray area, respectively.

From Fig. \ref{phase}(a), one can see that around $\theta/\pi \simeq$ 0 and 2, topological insulator phase with edge states within subband gap dominates for most of the $A$ values. Moreover, around $\theta/\pi \simeq$ 1 the finite-energy edges states associated with topologically nontrivial phases penetrate into the subband bulk states except for particular values $A\simeq$ 3 and 7. In these values of $A$, the finite-energy edge states completely reside within the main gap and, subsequently, the hybridized Floquet topological metal phase vanishes. Furthermore, for $\theta/\pi \simeq$ 0.5 and 1.5 with $A\simeq$ 4 and 9 the finite-energy edge states lie in the main gap emerging topological insulator phase.

As shown in Fig. \ref{phase}(b), the nontrivial topological phase associated with zero-energy edge states can be found for weak $A$ independent of $\theta$ values. But for intermediate and strong $A$ with $\theta/\pi \simeq$ 0.5 and 1.5 trivial insulator is dominated. Whereas for $\theta/\pi \simeq$ 0, 1, and 2 the phase changes from topological insulator to trivial one successively as a function of $A$.

\section*{Symmetric ladder case}\label{s5}

We consider symmetric ladder case where the dimerization pattern and lattice spacings of upper leg are the same as those for the lower leg as shown in Fig. \ref{fig1}(b). So, using Eq. (\ref{e6}), the hoppings of this case can be rewritten as
\begin{eqnarray}\label{e200}
\tilde{t}_1&=&\tilde{t}_2=t_1J_0[A_{x}(a_0-b_0)], \nonumber \\
\tilde{t}_3&=&\tilde{t}_4=t_3J_0[A_{y}c_0], \nonumber \\
\tilde{t_1^{\prime}}&=&\tilde{t_2^{\prime}}=t^{\prime}_1J_0[A_{x}b_0].
\end{eqnarray}
From the above equations, one finds that the horizontal (vertical) hoppings are affected only by x (y)-component of vector potential independent of $\phi$ due to rectangular symmetry of the lattice. This means that the circularly polarized light ($A_x=A_y=A$) can act as two independent linearly polarized fields in both directions. For this case, the Hamiltonian (\ref{e9}) commutes with exchange operator, $[\Upsilon, h_F(k)]=0$, and can be brought into a block diagonal form by the unitary matrix (\ref{e027}) as
\begin{eqnarray}\label{e18}
\tilde{h}_F=U_1 h_F(k) U_1^{-1}=\left(\begin{array}{c c}
 h_2&0\\
0 & -h_2
 \end{array}\right),
  \end{eqnarray}
  where
 \begin{eqnarray}\label{e19}
h_2=\left(\begin{array}{c c}
 \tilde{t}_{3}&\tilde{t}_1+\tilde{t}_1 ^\prime e^{ik}\\
\tilde{t}_1+\tilde{t}_1 ^\prime e^{-ik} & \tilde{t}_3
\end{array}\right).
\end{eqnarray}
This indicates that the existence of exchange symmetry will prevent the hybridization of edge states with the bulk states because the coupling block, $h_{cou}$, is zero. Likewise, the zero-energy edge states will be suppressed. Therefore, one may anticipate that the spectra of each block overlap with those of the other block so that the finite-energy edge states of a subsystem cross through bulk states of the other one without hybridization.

In the symmetric ladder model, there is time-reversal symmetry defined by $\mathcal{T}_i \tilde{h}_F(k)\mathcal{T}_i=\tilde{h}^\star_F(-k)$ (with i=1,2) where $\mathcal{T}_1=\sigma_0\otimes\sigma_0 \mathcal{K}$ and $\mathcal{T}_2=\sigma_z\otimes\sigma_0 \mathcal{K}$. Since the system has two particle-hole operators  $\mathcal{P}_1=\sigma_x\otimes\sigma_0\mathcal{K}$ and $\mathcal{P}_2=\sigma_y\otimes\sigma_0\mathcal{K}$ satisfying $\mathcal{P}_i \tilde{h}_F(k)\mathcal{P}_i=-\tilde{h}^\star_F(-k)$, the corresponding chiral operators fulfilling the sublattice symmetry $\mathcal{C}_i\tilde{h}_F(k)\mathcal{C}_i=-\tilde{h}_F(k)$ can be determined as
\begin{eqnarray}\label{e20}
\mathcal{C}_1 &=& \sigma_x \otimes \sigma_0, \\ \nonumber
\mathcal{C}_2 &=& \sigma_y \otimes \sigma_0.
\end{eqnarray}
Also, the Hamiltonian (\ref{e18}) has two inversion symmetry operators as
\begin{eqnarray}\label{e21}
\Pi_1 &=& \sigma_0 \otimes \sigma_x,  \\\nonumber
\Pi_2 &=& \sigma_z \otimes \sigma_x.
\end{eqnarray}
According to the above-mentioned symmetry statements, the symmetry operators exhibit the features that $\mathcal{T}^2=1$,  $\mathcal{P}^2=1$, and $\mathcal{C}^2=1$. Therefore, the symmetry class is still BDI \cite{Class10,AZClass1,AZClass3,AZClass4}. However, the diagonal blocks do not fall in BDI class.

We can obtain the eigenvalues of the model by diagonalizing Hamiltonian  (\ref{e18}) yielding
\begin{equation}\label{e22}
E=\pm \tilde{t}_3\pm\sqrt{\tilde{t}_1+\tilde{t}_1^\prime +2\tilde{t}_1\tilde{t}_1^\prime cos(k)}.
\end{equation}
Note, this energy spectrum is reminiscent of the spectrum of SSH model with the additional term $\tilde{t}_{3}$ which can be tuned by externally applied light. Such additional term acts like Zeeman field splitting the energy levels of SSH chain \cite{SSHZeeman}. When the vertical hopping $\tilde{t}_3=0$, the model reduces to two decoupled SSH chains with two-fold degenerate bulk states and the two dispersive finite-energy edge states convert to flat zero-energy edge states with four-fold degeneracy.

\begin{figure*}[t]
  \centering
    \includegraphics[width=1\linewidth]{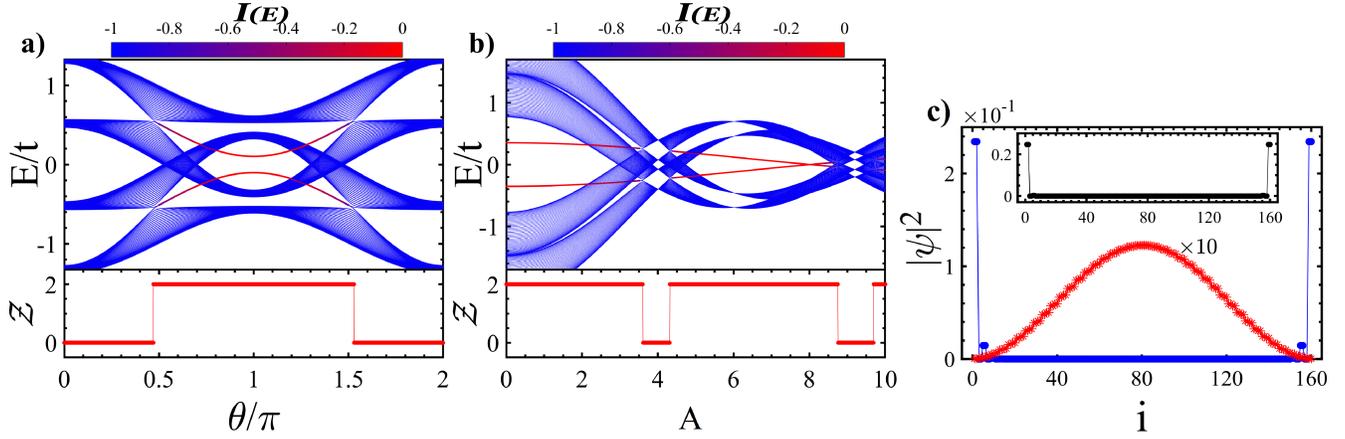}
    \caption{(Color online) Dependence of quasi-energy spectrum and its relevant topological invariant $\mathcal{Z}$ on (a) $\theta$ with $A = 5$ and on (b) $A$ with $\theta/\pi$ =0.8. The colors in the energy spectrum represent IPR of the wave function localization. (c) The probability distribution of energy states; Main panel: bulk states (the red curve with aster symbol) and the edge states within bulk states (the blue curve with circle symbol). Inset: the localized edge states within band gap. Here, $b_0=0.6$ and $c_0=0.3$.}
\label{fig2}
\end{figure*}

\begin{figure}[t!]
\centerline{\includegraphics[width=7cm]{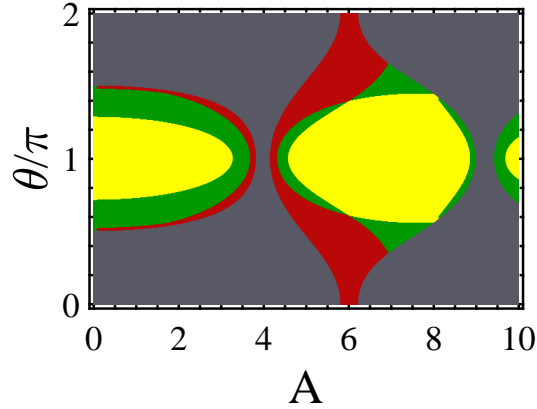}}
\caption{(Color online) Topological phase diagram in the plane ($A, \theta/\pi$) for symmetric ladder case. Yellow and red regions show topological insulator phase having $\mathcal{Z}$=2 with edge states within the main gap and subband gap, respectively. Green and gray regions indicate the Floquet topological metal state ($\mathcal{Z}$=2) and normal insulator ($\mathcal{Z}$=0). The parameters are $b_0=0.6$ and $c_0=0.3$.}
\label{fig3}
\end{figure}

As already mentioned above, for the present model, applying the circularly polarized light modifies the hoppings in the x-direction and y-direction independently. However, the topological phase transition again occurs at $k=0$ and $k=\pi$. So, by plugging Eq. (\ref{e200}) into Eq. (\ref{e12}), the gap closure/reopening conditions reduce as
\begin{eqnarray}\label{e121}
\tilde{t}_1 =-e^{ik} \tilde{t}_1^\prime.
\end{eqnarray}
Note that this relation which depends only on the horizontal hoppings is similar to the topological phase transition condition of original SSH model. So, the vertical hoppings have no effect on the topological phase transition points taking place at $\theta/\pi=$0.5 and $\theta/\pi=$1.5 in the static limit \cite{SSH,SSH1}. However, the energy levels at which gap closes are not zero and will be shifted by $\tilde{t}_3$ [see also Eq. (\ref{e22})] which is in contrast to the original SSH model [see Fig. \ref{fig2}(a)].

In Figs. \ref{fig2}(a) and \ref{fig2}(b), we have plotted the quasi-energy spectra along with bulk topological invariants versus $\theta/\pi$ and $A$, respectively. As already discussed, there are no zero-energy edge states and also the energy levels of finite-energy edge states change as functions of $\theta/\pi$ and $A$. From both figures, one can see that the finite-energy edge states penetrate into the bulk states and leave their band gap without occurring topological phase transition. Unlike the asymmetric ladder case, interestingly, due to presence of the exchange symmetry, the finite-energy edge states appear in the bulk states without hybridization \cite{Tm1,Tm2} resulting in Floquet topological metal phase.

Also, the probability distribution in terms of unit cell index along the ladder is shown in Fig. \ref{fig2}(c) for bulk states and finite-energy edge states in the bulk and in the gap. The finite-energy edge states remain localized within bulk and gapped states as indicated by the curves with blue "circle" symbols in the main panel and black "circle" symbols in the inset, respectively. Whereas the bulk states themselves exhibit extended feature [see the red curves with "star" symbols in the main panel].

In Fig. \ref{fig3}, the topological phase diagram is depicted in the ($A, \theta/\pi$)-plane. Also, we have distinguished the topological phases with edge states in the gap of subbands and in the main gap by red and yellow colors, respectively. The Floquet topological metal phase and trivial insulator are indicated by green and gray colors. Except for certain values of $A$, for $\theta/\pi$ around 1 the topological insulator with edge states in the main gap is dominated. By going away from $\theta/\pi\simeq $ 1 and approaching $\theta/\pi \simeq$ 0, 2 the Floquet topological metal, the topological insulator with edge states in the subband gap, and trivial insulator take place for weak and intermediate $A$. If $A$ is strong enough, the region corresponding to topological insulator containing edge states in the subband gap vanishes. This trend is due to the decrease in energy of the finite-energy edge states as $A$ increases [see Fig. \ref{fig2}(b)].

\begin{figure*}[t!]
  \centering
    \includegraphics[width=0.7\linewidth]{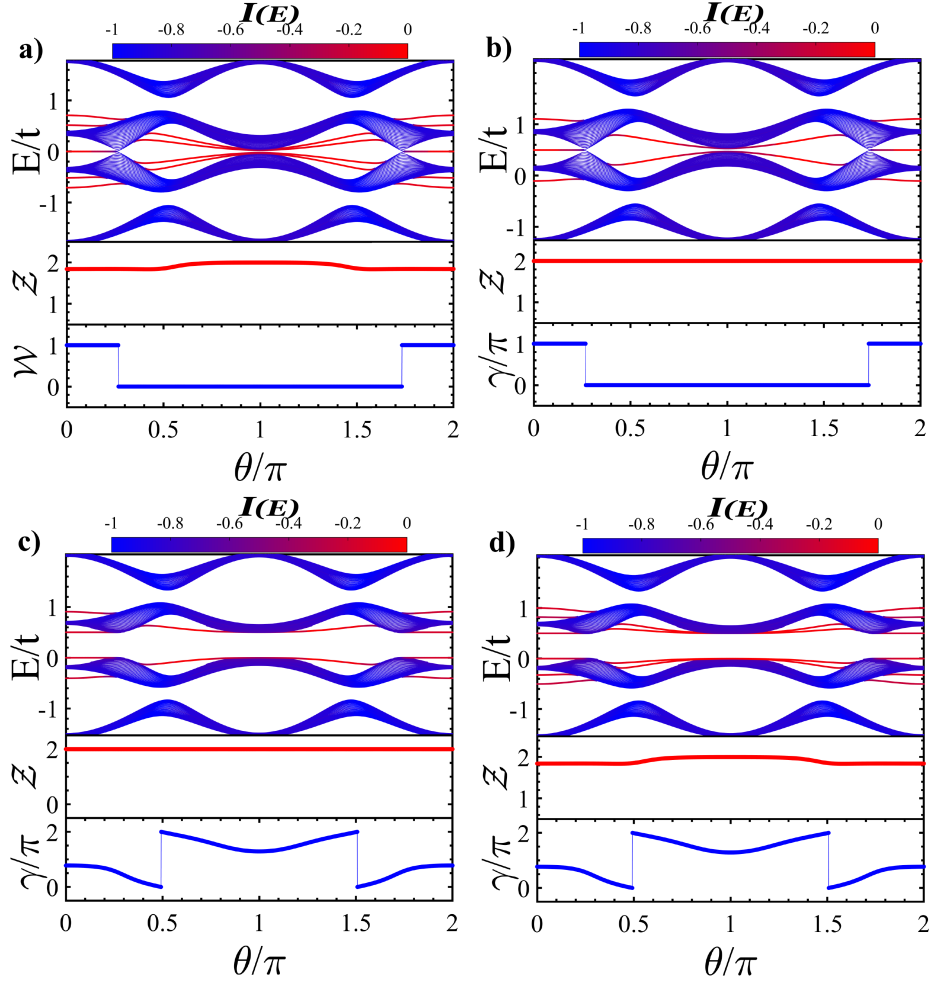}
    \caption{(Color online) Quasi-energy spectrum and the related topological invariants of the asymmetric ladder case exposed to the circularly polarized laser field as function of $\theta/\pi$ for (a) $\phi=\pi/4$ with the broken inversion symmetries of both diagonal blocks and  whole  Hamiltonian, (b) $\phi=\pi/2$ in the presence of $H^\prime$ with the broken chiral symmetry and preserved inversion symmetries of whole Hamiltonian and block $h_1$, (c) $\phi=\pi/2$ in the presence of $H^{''}$  with the broken inversion symmetry of whole Hamiltonian and preserved inversion symmetry of block $h_1$, and (d) $\phi=\pi/4$ in the presence of $H^{''}$ with the broken inversion symmetries of whole Hamiltonian and block $h_1$ as well as chiral symmetry. Here, the parameters are the same as Fig. \ref{fig4} and $V=t/2$.}
    \label{fig6}
\end{figure*}

\section*{Stability Of Edge States}\label{s6}

Now, we examine the stability of topological phases and demonstrate that which symmetry is responsible for the appearance of edge states. To do so, we consider the asymmetric ladder case subjected to a circularly polarized field in order to have maximum number of symmetry protected edge states, including zero- and finite-energy edge states.

Before illustrating the stability of topological edge states against perturbations like on-site potentials, we discuss about the effect of circular polarization of light with $\phi \ne n\pi/2$ on the topological characteristics of asymmetric ladder. According to Eq. (\ref{e60}), for $\phi \ne n\pi/2$, the two hoppings along the rungs are not equal, $\tilde{t}_3 \ne \tilde{t}_4$, resulting in the breaking of the inversion symmetry of the diagonal blocks ($h_1, -h_1$). This subsequently breaks the inversion symmetry of whole Hamiltonian as well. Consequently, the lack of inversion symmetry in the block $h_1$ gaps out the gapless finite-energy edge states lifting their degeneracy so that their relevant invariant $\mathcal{Z}$ takes continuous values as shown in Fig. \ref{fig6}(a). But despite the absence of inversion symmetry in the whole Hamiltonian, one can see that the zero-energy edge states and their relevant invariant $\mathcal{W}$ remain topologically nontrivial because of preserving the chiral symmetry.

In what follows, we assume $\phi=\pi/2$, otherwise specified. We add the on-site potential
\begin{equation}\label{e29}
H^\prime =V \sum _n A^{\dagger}_{uj}A_{uj}+B^{\dagger}_{uj}B_{uj}+A^{\dagger}_{lj}A_{lj}+B^{\dagger}_{lj}B_{lj},
\end{equation}
to the Hamiltonian (\ref{e7}) with $V$ being the amplitude of on-site potential. Moreover, the existence of $H^\prime$ breaks the chiral symmetry of whole Hamiltonian and shifts the energy levels as depicted in Fig. \ref{fig6}(b). But because of preserving the inversion symmetry of whole Hamiltonian, the multi-band Zak phase (\ref{e39}) can be employed as the topological invariant to characterize the topology of midgap edge states near the zero energy taking quantized values [see Fig. \ref{fig6}(b)]. Also, the inversion symmetry of diagonal block is preserved and the finite-energy edge states remain intact. On the other hand, we add the on-site potential of the form
\begin{equation}\label{e291}
H^{''} =V \sum _n A^{\dagger}_{uj}A_{uj} + B^{\dagger}_{lj}B_{lj},
\end{equation}
to Hamiltonian (\ref{e7}). This perturbation breaks both the chiral symmetry and the inversion symmetry of whole Hamiltonian while it preserves the inversion symmetry of blocks ($h_1, -h_1$). This means that the topological properties cannot transferred from the diagonal blocks to the full Hamiltonian. As shown in Fig. \ref{fig6}(c), the topology of zero-energy edge states is destroyed, however, the finite-energy edge states remain degenerate and nontrivial. As a result, the zero-energy edge states are protected by either the chiral symmetry or inversion of whole Hamiltonian.

Finally, We add the on-site potential $H^{''}$ to the system that is exposed to the circularly polarized field with $\phi=\pi/4$. In such situation, the chiral symmetry and inversion symmetry of whole Hamiltonian as well as the inversion symmetry of blocks ($h_1, -h_1$) will be broken. In Fig. \ref{fig6}(d), we have plotted the band structure illustrating that the finite- and zero-energy edge states are gapped with trivial values of their topological numbers. Consequently, the inversion symmetry of block $h_1$ is the fundamental symmetry protecting finite-energy edge states.

\section*{Summary} \label{s7}

We studied topological features of the two-leg SSH ladder periodically driven by circularly polarized light uncovering the role of lattice geometry. We considered asymmetric and symmetric ladders whose legs, respectively, have different and identical patterns of dimerization as well as lattice spacings. We found that there exist zero- and finite-energy edge states in the asymmetric ladder case, whereas the symmetric ladder hosts only the finite-energy ones. In both ladder models, the finite-energy edge states can leave from the gap of subbands and enter into the gap between the upper valence and lower conduction bands by crossing through the bulk states of subbands depending on the dimerization strength and driving amplitude. For asymmetric ladder, when the finite-energy edge states are within the bulk ones, due to the absence of exchange symmetry, these two types of states having the same energy and quantum number would hybridize together providing the hybridized Floquet topological metal states. Such new topological states are no longer localized. In contrast, for symmetric ladder case, the presence of exchange symmetry prevents hybridization between the finite-energy edge and bulk states establishing the Floquet topological metal phase with localized edge states. We also obtained the topological phase diagram that in addition of the two above-mentioned topological phases it contains a usual topological insulator and ordinary insulator. Furthermore, based on underlying symmetries of the system, we introduced relevant topological invariants to show the topology of the edge states. By involving symmetry breaking perturbations, we demonstrated that the finite-energy edge states are protected by the inversion symmetry of the diagonal blocks of Hamiltonian. But, the zero-energy edge states are protected by either the inversion or chiral symmetry of whole Hamiltonian. Moreover, we obtained an analytical formula for winding number to show the topology of zero-energy edge states when the chiral symmetry exists.

Finally, we note interestingly that interleg and intraleg hopping, respectively, can play the same roles as realistic Zeeman field and spin-orbit coupling effectively in our spinless model. So, such ingredients may not be necessary for quasi-1D systems \cite{doubleLadder}, unlike the topological 1D systems, to establish topological phases. This provides an alternative route to simulate Zeeman field and spin-orbit interaction in the absence of spin degree of freedom by engineering the existing degrees of freedom, for example, sublattice space. Furthermore, current experimental status can provide a possibility to realize two-leg ladder composed of coupled SSH chains \cite{ladder} and can manifest the topological signatures employing density and momentum-distribution measurements \cite{ExperLadder}. Also, the possible topological states can be recognized by using spatially resolved radio-frequency spectroscopy from the local density of states \cite{Expermesur}.

{\it Note added.} After completing the present study, we became aware that the Floquet topological metal phase has been investigated in Ref. \cite{FloMetal,FloSemiMetal}. In these works although the edge states can have the same energy as bulk states but, unlike our case, they are left isolated inside the band gap.


\section*{Author contributions}
M. V. Hosseini conceived the idea of the research and directed the project. All authors developed the research conceptions, analysed, discussed the obtained results, and wrote the paper. M. Jangjan performed the calculations.

\section*{Additional information}
\textbf{Competing financial interests:} The authors declare no competing financial and non-financial interests.

\end{document}